\documentclass[aps,preprint]{revtex4}%
\usepackage{amsfonts}
\usepackage{amsmath}
\usepackage{amssymb}
\usepackage{graphicx}%
\providecommand{\U}[1]{\protect\rule{.1in}{.1in}}

\begin{document}
\title[ ]{Classical Electromagnetic Derivation of the Bohr Spectrum}
\author{Timothy H. Boyer}
\affiliation{Department of Physics, City College of the City University of New York, New
York, New York 10031}

\begin{abstract}
Classical electrodynamics including Lorentz-invariant classical
electromagnetic zero-point radiation leads to a ground state and resonant
excited states for a charged particle in a Coulomb potential. \ These resonant
states correspond to integer values of the action variables analogous to those
appearing in the Bohr theory of the hydrogen atom. \ Thus there is a purely
classical electromagnetic derivation of the Bohr energy levels. \ According to
classical electromagnetic theory, it is not that there are states where the
charged particle does not radiate, but rather in these states there is a
resonant balance between radiation emission and absorption. The work on
\textit{classical zero-point radiation} reported here is a continuation of the
analysis published in 1975, but with the addition of the ideas of
\textit{relativity} and \textit{resonance} between the charge-particle orbit
and classical zero-point radiation. \ 

\end{abstract}
\maketitle

\section{Introduction}

\subsection{Historical Commentary}

In 1975, a survey article\cite{B1975} considered the progress made in
analyzing classical electrodynamics with random classical zero-point radiation
included. \ The mechanical system of choice was the simple harmonic
oscillator, the system first treated by Marshall\cite{Marshall2} in 1963.
\ Indeed one could easily show\cite{B1975} that the ground state and the
behavior in thermal radiation for a small oscillator was strongly reminiscent
of the quantum oscillator. \ Indeed, for free fields and harmonic oscillator
systems\cite{B1975b} the average values of Casimir forces and van der Waals
forces were the same as those given by quantum theory. \ It was assumed at
that time that all classical potentials were possible choices when discussing
the motion of a point charge. \ However, it was not long before it was
found\cite{B1976} that a small \textit{nonlinear} potential would scatter
Lorentz-invariant zero-point radiation toward the Rayleigh-Jeans spectrum.
\ Any nonlinear potential, such as a nonlinear oscillator potential, scattered
radiation away from the zero-point form. \ At this time, some physicists who
had been initially enthusiastic about including classical zero-point radiation
in classical electromagnetism turned against it. \ 

It was during this extended period of no progress, that Cole and
Zou\cite{Cole2003} came up with a computer simulation showing that a point
charge in a Coulomb potential in classical zero-point radiation gave a
probability distribution something like the Schroedinger ground state. \ In
subsequent years, Cole\cite{Cole2018} continued his work on simulations and
noted certain resonances for the Coulomb potential. \ Also, Batelaan and
Huang\cite{Batelaan} noted that there was some resonant-like behavior when a
radiation pulse fell on a point charge in a harmonic oscillator potential. \ 

However, other computer simulations seemed to cast doubt on the favorable
computer simulations. \ Ionization for the charged particle in a Coulomb
potential was claimed.\cite{Niewenhuizen2015} \ And the classical
electromagnetic point of view seemed to fail various experimental
observations. \ The idea of classical zero-point radiation was seen as an
interesting but failed effort. \ 

At present, the ideas of classical electrodynamics including random classical
zero-point radiation are in a revival. \ However, there are some important
changes from the point of view of 1975. \ 1) No longer are all potentials
$V\left(  r\right)  $ considered, but rather only \textit{relativistic} or
approximately relativistic potentials. \ Thus purely electromagnetic systems
or the limit of a point harmonic oscillator are allowed as relativistic or
approximately relativistic systems. \ 2) The gain of energy of a system from
random classical zero-point radiation must involve a \textit{resonance}
between the charged electromagnetic system and the zero-point radiation where
the finite extent of the charged system needs to be taken into account. \ The
harmonic oscillator is only approximately relativistic in the small oscillator
limit. \ On the other hand, the Coulomb potential is fully within relativistic
electrodynamics and needs no restriction. \ 

With these restrictions on allowed systems, one is able to discuss not only
the ground state of the harmonic oscillator but also its excited
states.\cite{B2026} \ The classical calculation agrees with the quantum
harmonic oscillator regarding its excited states. \ The Coulomb potential also
allows the treatment of excited states in agreement with Bohr's results for
the angular momentum, energy, frequency, and radius, as we show in the present
article. \ Furthermore, Cole's further simulation work\cite{Cole2026} seems to
show no evidence of ionization for the Coulomb potential. \ 

\subsection{Summary of the Present Article}

This present article assumes classical electromagnetic theory with the
boundary condition on the source-independent incoming radiation assumed to be
a Lorentz-invariant spectrum of random classical radiation. \ The scale of the
radiation is chosen so as to give the experimentally observed Casimir forces
between two conducting parallel plates. \ 

A charge in a Coulomb potential is treated in action-angle variables.
\ However, the numerical value of the action is not determined by the
mechanical motion. \ In this article, we calculate how the numerical value of
the action found in the zero-point radiation (where it is the same for all
radiation frequencies) is transferred to the action of the charge in the
Coulomb potential for both the ground state and the excited states. \ In the
nonrelativistic limit, we find the Bohr energy levels of 1913.

\section{Classical Zero-Point Radiation}

\subsection{Casimir Forces}

In 1948, Casimir\cite{Casimir} noted that the zero-point radiation of quantum
electrodynamics might lead to forces between conduction parallel plates.
\ This suggestion was confirmed by experimental work.\cite{Casimir2} \ The
experimental measurements, of the magnitude and distance-dependence of the
force, do not give any insight as to whether these are due to quantum or
classical electrodynamics. The force can be calculated in classical
electrodynamics involving the idea of random classical electromagnetic
zero-point radiation. \ In order to be Lorentz-invariant, the spectrum must
have one over-all constant for the average action. \ The average energy is
\begin{equation}
U^{zp}\left(  \omega\right)  =\left(  1/2\right)  \hbar\omega\label{zpr1}%
\end{equation}
where $\hbar$ is a constant taking the same numerical value as Planck's
constant.\cite{Planck} \ 

\subsection{Classical Zero-Point Radiation}

The source-free classical zero-point radiation in a very large
\textit{spherical} cavity of radius $\mathsf{R}$ can be written
as\cite{Jackson}%

\begin{align}
\mathbf{E}(\mathbf{r,}t)  &  =\operatorname{Re}\sum\nolimits_{n=1}^{\infty
}\sum\nolimits_{l=1}^{\infty}\sum\nolimits_{m=-l}^{m=l}\left\{  \exp\left[
i\left(  -k_{nl}^{M}ct+\theta_{nlm}^{M}\right)  \right]  \left[  ia_{nlm}%
^{M}j_{l}\left(  k_{nlm}^{M}r\right)  \mathbf{X}_{l,m}\left(  \theta
,\phi\right)  \right]  \right. \nonumber\\
&  \left.  +\exp\left[  i\left(  -k_{nlm}^{E}ct+\theta_{nlm}^{E}\right)
\right]  \left[  a_{nlm}^{E}/\left(  -ik_{nlm}^{E}\right)  \right]
\nabla\times\left[  j_{l}\left(  k_{nlm}^{E}r\right)  \mathbf{X}_{lm}\left(
\theta,\phi\right)  \right]  \right\}  , \label{Ezprt}%
\end{align}
and%

\begin{align}
\mathbf{B}(\mathbf{r,}t)  &  =\operatorname{Re}\sum\nolimits_{n=1}^{\infty
}\sum\nolimits_{l=1}^{\infty}\sum\nolimits_{m=-l}^{m=l}\left\{  \exp\left[
i\left(  -k_{nlm}^{E}ct+\theta_{nlm}^{E}\right)  \right]  \left[  ia_{nlm}%
^{E}j_{l}\left(  k_{nlm}^{E}r\right)  \mathbf{X}_{l,m}\left(  \theta
,\phi\right)  \right]  \right. \nonumber\\
&  \left.  +\exp\left[  i\left(  -k_{nlm}^{M}ct+\theta_{nlm}^{M}\right)
\right]  \left[  a_{nlm}^{M}/\left(  ik_{nlm}^{M}\right)  \right]
\nabla\times\left[  j_{l}\left(  k_{nlm}^{M}r\right)  \mathbf{X}_{lm}\left(
\theta,\phi\right)  \right]  \right\}  , \label{Bzprt}%
\end{align}
where $a_{nlm}^{E}$ and $a_{nlm}^{M}$ are the amplitudes of the electric and
magnetic modes, $j_{l}$ is the spherical Bessel function of order $l$,
$\mathbf{X}_{l,m}\left(  \theta,\phi\right)  $ is the vector spherical
harmonic, and the phases $\theta_{nlm}^{E}$ and $\theta_{nlm}^{M}$\ are the
random phases distributed uniformly on $(0,2\pi]$ and independently for each
mode.\cite{EH}\cite{Rice} \ The amplitudes of the spherical standing waves
correspond to the spectrum $U^{zp}\left(  c\mathbf{k}\right)  =\frac{1}%
{2}\hbar ck$ given in Eq. (\ref{zpr1}), where the angular frequency
$\omega=ck=c\left\vert \mathbf{k}\right\vert $, and the randomness is
introduced by the random phases for the electric and magnetic modes
$\theta_{nlm}^{E}$ and $\theta_{nlm}^{M}$. \ The representation of the
rotation group refers to the integers $l$ and $m$ which are unrelated to the
frequency $\omega$ of the normal mode.

Here we have the vector spherical harmonic\cite{Jackson2}%
\begin{equation}
\mathbf{X}_{lm}\left(  \theta,\phi\right)  =\frac{\mathbf{L}Y_{lm}\left(
\theta,\phi\right)  }{\sqrt{l\left(  l+1\right)  }}=\frac{\mathbf{r\times
\nabla}Y_{l,m}\left(  \theta,\phi\right)  }{i\sqrt{l\left(  l+1\right)  }},
\end{equation}
where the spherical harmonics are given by\cite{Jackson3}%
\begin{align}
Y_{lm}\left(  \theta=\pi/2,\phi\right)   &  =\left[  \sqrt{\frac{\left(
2l+1\right)  }{4\pi}\frac{\left(  l-m\right)  !}{\left(  l+m\right)  !}}%
P_{l}^{m}\left(  \cos\theta\right)  \right]  _{\theta=\pi/2}\exp\left[
im\phi\right] \nonumber\\
&  =Y_{lm}\left(  \pi/2,0\right)  \left\{  \cos\left[  m\phi\right]
+i\sin\left[  m\phi\right]  \right\}  ,
\end{align}
and the only complex part is the exponential behavior in $\phi$, $\exp\left[
im\phi\right]  $. \ 

For random radiation, we have the scale given by\cite{Disguised}%

\begin{equation}
\left\vert a_{nlm}^{E}\right\vert ^{2}=\frac{16\pi\left(  k_{nl}^{E}\right)
^{2}}{\mathsf{R}}\left[  U_{nlm}\left(  k_{nlm}^{E}\right)  \right]  .
\label{aEnlm2}%
\end{equation}

The situation for \textit{magnetic} modes $a_{nlm}^{M}$ is exactly analogous.
\ Magnetic radiation modes contribute for a charged particle in a circular
orbit beginning at $l=2$ with a magnitude comparable to the \textit{electric}
multipole of order $l=3$.

For the limit of large radius $\mathsf{R}$ of the enclosing sphere containing
standing waves, we have\cite{Disguised} $dk=\pi dn/\mathsf{R}$ and \
\begin{equation}
\sum\nolimits_{n=1}^{\infty}\rightarrow\int_{0}^{\infty}dn=\int_{0}^{\infty
}dk\frac{\mathsf{R}}{\pi}=\int_{0}^{\infty}d\omega\frac{\mathsf{R}}{\pi c}.\,
\end{equation}
If we use random phases for closely-space radiation modes labeled by $nlm$,
then the amplitude involves the average energy $U_{nlm}.$ \ The number of
normal modes per unit (angular) frequency per unit volume is $\omega
^{2}/\left(  \pi^{2}c^{3}\right)  $ which gives an energy per unit angular
frequency interval per unit volume $\left[  \omega^{2}/\left(  \pi^{2}%
c^{3}\right)  \right]  U\left(  \omega\right)  .$ \ 

\section{Relativistic\ Particle in the Coulomb potential in Classical
Mechanics}

The behavior of a relativistic charge $e$ in the Coulomb potential in
action-angle variables with no radiation being emitted is given
by\cite{Goldstein2}%
\begin{equation}
U\left(  J_{r},J_{\theta},J_{\phi}\right)  =M_{0}c^{2}\left(  1+\left[
\frac{e^{2}/c}{J_{r}+\sqrt{\left(  J_{\theta}+J_{\phi}\right)  ^{2}-\left(
e^{2}/c\right)  ^{2}}}\right]  ^{2}\right)  ^{-1/2}, \label{UJrJtheta}%
\end{equation}
where the action variables are connected with the coordinates in spherical
coordinates. \ This expression does not give any magnitudes for the adiabatic
invariants $J_{r},~J_{\theta},\ $or $J_{\phi}$. \ We must go outside pure
mechanics in order to find a source for any particular values for the action variables.

If we wish to remain within classical theory, we assume the presence of
Lorentz-invariant, random, classical electromagnetic zero-point radiation.
\ Then the whole system (the charged particle, Coulomb potential, and
classical zero-point radiation) can be regarded as being within classical
electrodynamics. \ The momentum of the charged particle satisfies the
differential equations \
\begin{align}
\frac{d\mathbf{p}_{e}}{dt} &  =-e\frac{\mathbf{r}_{e}}{r_{e}^{3}}%
+e\mathbf{E}\left[  \mathbf{r}_{e}(t),t\right]  +e\frac{\mathbf{v}%
_{e}\mathbf{(}t)}{c}\times\mathbf{B}\left[  \mathbf{r}_{e}(t),t\right]
\nonumber\\
&  =-e\frac{\mathbf{r}_{e}}{r_{e}^{3}}+e\left\{  \mathbf{-\nabla}\Phi\left[
\mathbf{r}_{e}(t),t\right]  -\frac{1}{c}\frac{\partial\mathbf{A}\left[
\mathbf{r}_{e}(t),t\right]  }{\partial t}\right\}  +e\frac{\mathbf{v}%
_{e}\mathbf{(}t)}{c}\times\left(  \mathbf{\nabla}\times\mathbf{A}\left[
\mathbf{r}_{e}(t),t\right]  \right)  ,\label{dpdt0}%
\end{align}
where%
\begin{equation}
\frac{d\mathbf{r}_{e}}{dt}=\frac{\mathbf{p}_{e}\mathbf{/}m}{\sqrt{1+\left[
p_{e}/\left(  mc\right)  \right]  ^{2}}}\text{ \ with \ }M_{0}\frac
{\mathbf{v}_{e}\left(  t\right)  }{\sqrt{1-v_{e}^{2}/c^{2}}}=\mathbf{p}%
_{e}\left(  t\right)  .
\end{equation}
The potential fields are given by%
\begin{equation}
\left(  \nabla^{2}-\frac{1}{c^{2}}\frac{\partial^{2}}{\partial t^{2}}\right)
\Phi\left(  \mathbf{r,}t\right)  =-4\pi e\delta^{3}\left[  \mathbf{r-r}%
_{e}(t)\right]
\end{equation}
and
\begin{equation}
\left(  \nabla^{2}-\frac{1}{c^{2}}\frac{\partial^{2}}{\partial t^{2}}\right)
\mathbf{A}\left(  \mathbf{r,}t\right)  =-4\pi e\frac{\mathbf{v}_{e}\left(
t\right)  }{c}\delta^{3}\left[  \mathbf{r-r}_{e}(t)\right]  .
\end{equation}
Note particularly the dependence of $\mathbf{E}$ and $\mathbf{B}$ on the
position $\mathbf{r}_{e}\left(  t\right)  $ of the charged particle.

\section{Gain of Energy for a Charge in a Circular Orbit in Zero-Point
Radiation}

\subsection{Equation of Motion for the Charge Driven by Random Radiation}

A charge $e$ in a \textit{circular} orbit in the $xy$-plane will experience a
force due to the $\phi$-component of the electric field $E_{\phi}^{E}%
(r_{e},\pi/2,\phi_{e}\left(  t\right)  ,t)$ at the angular location $\phi
_{e}\left(  t\right)  $ of the charge. \ Newton's second law gives%
\begin{equation}
Mr_{e}^{2}\frac{d^{2}\phi_{e}\left(  t\right)  }{dt^{2}}=r_{e}eE_{\phi}%
^{E}(\mathbf{r}_{e}\left(  t\right)  ,t), \label{Mr2dphi}%
\end{equation}
where $M=M_{0}\gamma$ is the relativistic particle mass which is assumed not
to change during the short time $\tau$. \ For the solution of this equation,
we use variation of parameters analogous to Born's work\cite{Born2} for the
linear harmonic oscillator. \ Now the \textit{source-free} solutions of the
homogeneous equation $Mr_{0}^{2}\ddot{\phi}=0$ for the circular orbit are
$\phi\left(  t\right)  =a$ and $\phi\left(  t\right)  =bt$ where $a$ and $b$
are constants. \ Then it follows that the particular solution $\phi_{p}(t)$ of
Eq. (\ref{Mr2dphi}) is obtained by variation of parameters as\cite{Greenberg}%

\begin{align}
Mr_{e}\phi_{p}(t) &  =-\int_{0}^{t^{\prime}=t}dt^{\prime}\left[  eE_{\phi}%
^{E}(\mathbf{r}_{e}\left(  t^{\prime}\right)  ,t^{\prime})\right]  t^{\prime
}+\int_{0}^{t^{\prime}=t}dt^{\prime}\left[  eE_{\phi}^{E}(\mathbf{r}%
_{e}\left(  t^{\prime}\right)  ,t^{\prime})\right]  t\nonumber\\
&  =\int_{0}^{t^{\prime}=t}dt^{\prime}\left[  eE_{\phi}^{E}(\mathbf{r}%
_{e}\left(  t^{\prime}\right)  ,t^{\prime})\right]  \left(  t-t^{\prime
}\right)  .\label{Mrphip}%
\end{align}
Note that the angular velocity $d\phi_{p}\left(  t\right)  /dt$ follows as%
\begin{align}
Mr_{e}\frac{d\phi_{p}(t)}{dt} &  =-\left[  eE_{\phi}^{E}(\mathbf{r}_{e}\left(
t\right)  ,t)\right]  t+\left[  eE_{\phi}^{E}(\mathbf{r}_{e}\left(  t\right)
,t)\right]  t\nonumber\\
&  +\int_{0}^{t^{\prime}=t}dt^{\prime}\left[  eE_{\phi}^{E}(\mathbf{r}%
_{e}\left(  t^{\prime}\right)  ,t^{\prime})\right]  \nonumber\\
&  =\int_{0}^{t^{\prime}=t}dt^{\prime}\left[  eE_{\phi}^{E}(\mathbf{r}%
_{e}\left(  t^{\prime}\right)  ,t^{\prime})\right]  ,\label{Mr0v}%
\end{align}
and indeed the equation of motion (\ref{Mr2dphi}) is satisfied, since when we
take the second derivative with respect to time $t$, we obtain \
\begin{equation}
Mr_{e}\frac{d^{2}\phi_{p}(t)}{dt^{2}}=\left[  eE_{\phi}^{E}(\mathbf{r}%
_{e}\left(  t\right)  ,t)\right]  .
\end{equation}
Thus the proposed solution in Eq. (\ref{Mrphip}) indeed satisfies the
differential equation (\ref{Mr2dphi}), and also satisfies the boundary
conditions $\phi_{p}\left(  0\right)  =0$ and $\dot{\phi}_{p}\left(  0\right)
=0$ for the circular orbit. \ 

\subsection{Energy Gain from Random Radiation}

The energy delivered by the random radiation during a short time interval
$\tau$ involving many orbital revolutions, but little change in the mechanical
angular momentum $J_{\phi}$ of the particle is
\begin{align}
W\left(  \tau\right)   &  =\int_{0}^{\tau}dt\left[  eE_{\phi}^{E}%
(\mathbf{r}_{e}\left(  t\right)  ,t)\right]  \left[  r_{e}\frac{d\phi_{p}%
(t)}{dt}\right] \nonumber\\
&  =\frac{1}{M}\int_{0}^{\tau}dt\left[  eE_{\phi}^{E}(\mathbf{r}_{e}\left(
t\right)  ,t)\right]  \left\{  \int_{0}^{t^{\prime}=t}dt^{\prime}\left[
eE_{\phi}^{E}(\mathbf{r}_{e}\left(  t^{\prime}\right)  ,t^{\prime})\right]
\right\} \nonumber\\
&  =\frac{e^{2}}{M}\int_{0}^{\tau}dt\int_{0}^{t^{\prime}=t}dt^{\prime}\left[
E_{\phi}^{E}(\mathbf{r}_{e}\left(  t\right)  ,t)\right]  \left[  E_{\phi}%
^{E}(\mathbf{r}_{e}\left(  t^{\prime}\right)  ,t^{\prime})\right]  ,
\label{WPtauP}%
\end{align}
where again $M=M_{0}\gamma$ is the relativistic mass. \ The integrand in Eq.
(\ref{WPtauP}) is \textit{symmetric} under interchange of $t$ and $t^{\prime}%
$. \ \ But then we can carry through Born's symmetrizing. \ Thus, the double
integral over the isosceles triangular region for $t$ and $t^{\prime}$ in Eq.
(\ref{WPtauP}) is the same as obtained by integrating first in $t$ from
$t^{\prime}$ to $\tau$, and then in $t^{\prime}$ from $0$ to $\tau$ as%

\begin{align}
&  \int_{0}^{\tau}dt\int_{0}^{t^{\prime}=t}dt^{\prime}\left[  E_{\phi}%
^{E}(\mathbf{r}_{e}\left(  t\right)  ,t)\right]  \left[  E_{\phi}%
^{E}(\mathbf{r}_{e}\left(  t^{\prime}\right)  ,t^{\prime})\right] \nonumber\\
&  =\int_{0}^{\tau}dt^{\prime}\int_{t=t^{\prime}}^{t=\tau}dt\left[  E_{\phi
}^{E}(\mathbf{r}_{e}\left(  t\right)  ,t)\right]  \left[  E_{\phi}%
^{E}(\mathbf{r}_{e}\left(  t^{\prime}\right)  ,t^{\prime})\right]  .
\end{align}
However, we may interchange the prime and unprime labels and add half the
expressions to obtain%
\begin{equation}
W\left(  \tau\right)  =\frac{e^{2}}{2M}\int_{0}^{t=\tau}dt\int_{0}^{t^{\prime
}=\tau}dt^{\prime}\left[  E_{\phi}^{E}(\mathbf{r}_{e}\left(  t\right)
,t)\right]  \left[  E_{\phi}^{E}(\mathbf{r}_{e}\left(  t^{\prime}\right)
,t^{\prime})\right]  , \label{Wtaue2}%
\end{equation}
where now both time integrals are from $0$ to $\tau$. \ 

\subsection{Averaging Over the Random Phases of the Zero-Point Radiation}

Now we need the \textit{average} energy absorbed in the short time $\tau$
involving many passages over the charged-particle orbit. \ Thus, equation
(\ref{Wtaue2}) becomes%
\begin{equation}
\left\langle W\left(  \tau\right)  \right\rangle _{\theta^{E}}=\frac{e^{2}%
}{2M}\int_{0}^{\tau}dt\int_{0}^{\tau}dt^{\prime}\left\langle \left[  E_{\phi
}^{E}(\mathbf{r}_{e}\left(  t\right)  ,t)\right]  \left[  E_{\phi}%
^{E}(\mathbf{r}_{e}\left(  t^{\prime}\right)  ,t^{\prime})\right]
\right\rangle _{\theta^{E}}, \label{AvWtau}%
\end{equation}
where we average over the random phases $\theta^{E}$ of the driving radiation. \ 

The $\phi$-component of the electric field due to zero-point radiation can be
rewritten as
\begin{align}
\left[  E_{\phi}^{E}(\mathbf{r}_{e}\left(  t\right)  ,t)\right]   &
=\operatorname{Re}E_{\phi}^{E}\left(  \mathbf{r}_{e}\left(  t\right)
,0\right)  \exp\left[  -i\omega^{E}t+i\theta^{E}\right]  \nonumber\\
&  =E_{\phi}^{E}\left(  \mathbf{r}_{e}\left(  t\right)  ,0\right)  \cos\left[
\omega^{E}t-\theta^{E}\right]  .
\end{align}
Random radiation can be treated by introducing random phases for the waves.
\ However, there is already a phase in the problem, that of the charged
particle in its orbit. \ Thus we must be careful to preserve the randomness of
the radiation waves relative to the phase already present in the orbital
motion. \ In equation (\ref{AvWtau}), there are two phases corresponding to
the two different radiation waves at times $t$ and $t^{\prime}$ involved in
the integrals. \ However, if the two phases are different, the integrals over
many periods will give vanishing results. \ Thus the only contribution which
survives the integration is that where a radiation wave is matched with
itself. \ However, this radiation wave will still retain its randomness
compared with the phase of the orbiting charge. \ 

We are interested in averaging over the random phases $\theta_{n^{\prime
}l^{\prime}m^{\prime}}^{E}$ to obtain $\left\langle W\left(  \tau\right)
\right\rangle _{\theta^{E^{\prime}}}$ as in Eq. (\ref{AvWtau}). \ We require%
\begin{equation}
\left\langle \cos\left[  \theta_{nlm}^{E}\right]  \cos\left[  \theta
_{n^{\prime}l^{\prime}m^{\prime}}^{E}\right]  \right\rangle _{\theta
^{E^{\prime}}}=\left\langle \sin\left[  \theta_{nlm}^{E}\right]  \sin\left[
\theta_{n^{\prime}l^{\prime}m^{\prime}}^{E}\right]  \right\rangle
_{\theta^{E^{\prime}}}=\frac{1}{2}\delta_{nlm,n^{\prime}l^{\prime}m^{\prime}},
\end{equation}
and%
\begin{equation}
\left\langle \cos\left[  \theta_{nlm}^{E}\right]  \sin\left[  \theta
_{n^{\prime}l^{\prime}m^{\prime}}^{E}\right]  \right\rangle _{\theta
^{E^{\prime}}}=0.
\end{equation}

\ Then for the electric multipole radiation, averaging and then summing to
eliminate the $\delta_{nlm,n^{\prime}l^{\prime}m^{\prime}}$, we have%
\begin{align}
&  \left\langle \left[  E_{\phi}^{E}(\mathbf{r}_{e}\left(  t\right)
,t)\right]  \left[  E_{\phi}^{E}(\mathbf{r}_{e}\left(  t^{\prime}\right)
,t^{\prime})\right]  \right\rangle _{\theta^{E^{\prime}}}\nonumber\\
&  =\left\langle \sum\nolimits_{nlm}E_{\phi}^{E}\left(  \mathbf{r}_{e}\left(
t\right)  ,0\right)  \cos\left[  \omega^{E}t-\theta_{nlm}^{E}\right]
\sum\nolimits_{n^{\prime}l^{\prime}m^{\prime}}E_{\phi}^{E}\left(
\mathbf{r}_{e}\left(  t^{\prime}\right)  ,0\right)  \cos\left[  \omega
^{E}t^{\prime}-\theta_{n^{\prime}l^{\prime}m^{\prime}}^{E}\right]
\right\rangle _{\theta^{E^{\prime}}}\nonumber\\
&  =\sum\nolimits_{nlm}E_{nlm-\phi}^{E}(\mathbf{r}_{e}\left(  t\right)
,0)\sum\nolimits_{n^{\prime}l^{\prime}m^{\prime}}E_{n^{\prime}l^{\prime
}m^{\prime}-\phi}^{E}(\mathbf{r}_{e}\left(  t^{\prime}\right)  ,0)\cos\left[
\omega_{n^{\prime}l^{\prime}m^{\prime}}^{E}\left(  t-t^{\prime}\right)
\right]  \frac{1}{2}\delta_{nlm,n^{\prime}l^{\prime}m^{\prime}}\nonumber\\
&  =\frac{1}{2}\sum\nolimits_{nlm}\left[  E_{nlm-\phi}^{E}(\mathbf{r}%
_{e}\left(  t\right)  ,0)\right]  \left[  E_{nlm-\phi}^{E}(\mathbf{r}%
_{e}\left(  t^{\prime}\right)  ,0)\right]  \cos\left[  \omega_{nlm}^{E}\left(
t-t^{\prime}\right)  \right]  .
\end{align}
Then the average energy absorbed in Eq. (\ref{AvWtau}) is
\begin{align}
&  \left\langle W^{E}\left(  \tau\right)  \right\rangle _{\theta^{E^{\prime}}%
}\nonumber\\
&  =\frac{e^{2}}{2M}\int_{0}^{\tau}dt\int_{0}^{\tau}dt^{\prime}\left\langle
\left[  E_{\phi}^{E}(\mathbf{r}_{e}\left(  t\right)  ,t)\right]  \left[
E_{\phi}^{E}(\mathbf{r}_{e}\left(  t^{\prime}\right)  ,t^{\prime})\right]
\right\rangle \nonumber\\
&  =\frac{e^{2}}{2M}\int_{0}^{\tau}dt\int_{0}^{\tau}dt^{\prime}\frac{1}{2}%
\sum\nolimits_{nlm}\left[  E_{nlm-\phi}^{E}(\mathbf{r}_{e}\left(  t\right)
,0)\right]  \left[  E_{nlm-\phi}^{E}(\mathbf{r}_{e}\left(  t^{\prime}\right)
,0)\right]  \cos\left[  \omega_{nlm}^{E}\left(  t-t^{\prime}\right)  \right]
. \label{wtauq}%
\end{align}
Now we can use the expansion
\begin{equation}
\cos\left[  \omega_{nlm}^{E}\left(  t-t^{\prime}\right)  -\theta_{nlm}%
^{E}\right]  =\cos\left[  \omega_{nlm}^{E}t-\theta_{nlm}^{E}\right]
\cos\left[  \omega_{nlm}^{E}t^{\prime}\right]  +\sin\left[  \omega_{nlm}%
^{E}t-\theta_{nlm}^{E}\right]  \sin\left[  \omega_{nlm}^{E}t^{\prime}\right]
,
\end{equation}
and therefore can rewrite Eq. (\ref{wtauq}) as%
\begin{align}
\left\langle W^{E}\left(  \tau\right)  \right\rangle  &  =\frac{e^{2}}{4M}%
\sum\nolimits_{nlm}\left\{  \left[  \int_{0}^{\tau}dt\left[  E_{nlm-\phi}%
^{E}(\mathbf{r}_{e}\left(  t\right)  ,0)\right]  \cos\left[  \omega_{nlm}%
^{E}t\right]  \right]  ^{2}\right. \nonumber\\
&  +\left.  \left[  \int_{0}^{\tau}dt\left[  E_{nlm-\phi}^{E}(\mathbf{r}%
_{e}\left(  t\right)  ,0)\right]  \sin\left[  \omega_{nlm}^{E}t\right]
\right]  ^{2}\right\}  \label{avWtaue2}%
\end{align}

Now the charged particle is going around with frequency $\omega_{e-n}$ in a
circular orbit of radius $r_{e-n}$, so that its speed is $\omega_{e-n}%
r_{e-n}=v_{e-n}$. \ Accordingly, the first integral in the \textit{electric}
multipole field in Eq. (\ref{avWtaue2}) requires the expansion%

\begin{equation}
E_{\phi}^{E}\left[  \mathbf{r}_{e}\left(  t\right)  ,0\right]  =\sum
\nolimits_{n^{\prime}lm}^{\infty}E_{n^{\prime}lm-\phi}^{E}\left(
k_{n^{\prime}lm}^{E}r_{e}\right)  \cos\left[  \omega_{e-n}t\right]  .
\end{equation}
Then including the time behavior of the electric field, we have%
\begin{align}
&  \int_{0}^{\tau}dt\left[  E_{n^{\prime}lm-\phi}^{E}(\mathbf{r}_{e}\left(
t\right)  ,0)\right]  \cos\left[  \omega_{n^{\prime}lm}^{E}t\right]
\nonumber\\
&  =\int_{0}^{\tau}dt\left[  \sum\nolimits_{n^{\prime}lm}^{\infty}%
E_{n^{\prime}lm-\phi}^{E}\left(  k_{n^{\prime}lm}^{E}r\right)  \cos\left[
\omega_{e-n}t\right]  \right]  \cos\left[  \omega_{n^{\prime}lm}^{E}t\right]
\nonumber\\
&  =\int_{0}^{\tau}dt\sum\nolimits_{n^{\prime}lm}^{\infty}E_{n^{\prime}%
lm-\phi}^{E}\left(  k_{n^{\prime}lm}^{E}r\right)  \frac{1}{2}\left\{
\cos\left[  \left(  \omega_{e-n}-\omega_{nlm}^{E}\right)  t\right]
+\cos\left[  \left(  \omega_{e-n}+\omega_{nlm}^{E}\right)  t\right]  \right\}
\nonumber\\
&  =\sum\nolimits_{n^{\prime}lm}^{\infty}E_{n^{\prime}lm-\phi}^{E}\left(
k_{n^{\prime}lm}^{E}r\right)  \frac{1}{2}\left\{  \frac{\sin\left[  \left(
\omega_{e-n}-\omega_{n^{\prime}lm}^{E}\right)  \tau\right]  }{\left(
\omega_{e-n}-\omega_{n^{\prime}lm}^{E}\right)  }+\frac{\sin\left[  \left(
\omega_{e-n}+\omega_{n^{\prime}lm}^{E}\right)  \tau\right]  }{\left(
\omega_{e-n}+\omega_{n^{\prime}lm}^{E}\right)  }\right\}  ,
\end{align}
where $\omega_{e-n}$ is the frequency of the charged particle in its orbit.
\ For a circular orbit, the expression $\left[  E_{nlm-\phi}^{E}(r_{e}%
,\pi/2,0,0)\right]  $ does not involve time $t$ and so can be taken outside
the time integral. \ 

For a very large spherical enclosure, the sum turns into an integral, and we
may use Eqs. (\ref{Ezprt}) and (\ref{aEnlm2}) giving
\begin{align}
&  \left\langle W_{l,m}^{E}\left(  \tau\right)  \right\rangle \nonumber\\
&  =\frac{e^{2}}{4M}\sum\nolimits_{nlm}\left[  E_{nlm-\phi}^{E}(r_{e}%
,\pi/2,0,0)\right]  ^{2}\left\{  \left[  \int_{0}^{\tau}dt\cos\left[  \left(
\omega_{e}-\omega_{n^{\prime}l^{\prime}m^{\prime}}^{E}\right)  t\right]
\right]  ^{2}\right. \nonumber\\
&  +\left.  \left[  \int_{0}^{\tau}dt\sin\left[  \left(  \omega_{e}%
-\omega_{n^{\prime}l^{\prime}m^{\prime}}^{E}\right)  t\right]  \right]
^{2}\right\} \nonumber\\
&  =\frac{e^{2}}{4M}\left(  \int_{0}^{\infty}d\omega^{E}\frac{\mathsf{R}}{\pi
c}\right)  \left[  m^{2}\left[  \frac{1}{x}\left(  \frac{d\left[
xj_{l}\left(  x\right)  \right]  }{d\left(  x\right)  }\right)  \right]
_{x=mv/c}Y_{lm}\frac{a_{lm}^{E}}{\sqrt{l\left(  l+1\right)  }}\right]
_{\theta=\pi/2,\phi=0}^{2}\nonumber\\
&  \times\left\{  \left[  \int_{0}^{\tau}dt\cos\left[  \left(  \omega
_{e}-\omega^{E}\right)  t\right]  \right]  ^{2}+\left[  \int_{0}^{\tau}%
dt\sin\left[  \left(  \omega_{e}-\omega^{E}\right)  t\right]  \right]
^{2}\right\} \nonumber\\
&  =\frac{e^{2}}{4M}\left(  \int_{0}^{\infty}d\omega^{E}\frac{\mathsf{R}}{\pi
c}\right)  m^{4}\left[  \frac{1}{x}\left(  \frac{d\left(  xj_{l}\left(
x\right)  \right)  }{dx}\right)  \right]  _{x=mv/c}^{2}\frac{\left\vert
Y_{lm}\right\vert ^{2}}{l\left(  l+1\right)  }\left[  \frac{16\pi\left(
\omega_{nl}^{E}\right)  ^{3}}{c^{2}\mathsf{R}}\left\langle J_{rad}\left(
\omega_{nl}^{E}\right)  \right\rangle \right] \nonumber\\
&  \times\left\{  \left[  \int_{0}^{\tau}dt\cos\left[  \left(  \omega
_{e}-\omega^{E}\right)  t\right]  \right]  ^{2}+\left[  \int_{0}^{\tau}%
dt\sin\left[  \left(  \omega_{e}-\omega^{E}\right)  t\right]  \right]
^{2}\right\} \nonumber\\
&  =\frac{4\pi e^{2}}{M}\left(  \int_{0}^{\infty}d\omega^{E}\frac{\mathsf{R}%
}{\pi c}\right)  m^{4}\left[  \frac{1}{x}\left(  \frac{d\left(  xj_{l}\left(
x\right)  \right)  }{dx}\right)  \right]  _{x=mv/c}^{2}\frac{\left\vert
Y_{lm}\right\vert ^{2}}{l\left(  l+1\right)  }\left[  \frac{\left(
\omega_{nl}^{E}\right)  ^{3}}{c^{2}\mathsf{R}}\left\langle J_{rad}\left(
\omega_{nl}^{E}\right)  \right\rangle \right] \nonumber\\
&  \times\left\{  \left[  \frac{\sin^{2}\left[  \left(  \omega_{e}-\omega
^{E}\right)  \tau\right]  }{\left(  \omega_{e}-\omega^{E}\right)  ^{2}%
}\right]  +\left[  \frac{\left(  1-\cos\left[  \left(  \omega_{e}-\omega
^{E}\right)  \tau\right]  \right)  ^{2}}{\left(  \omega_{e}-\omega^{E}\right)
^{2}}\right]  \right\}  `
\end{align}

For a single spherical wave mode $l,m$ at frequency $\omega^{E}$, we integrate
over $\omega^{E}$ as
\begin{equation}
\int_{0}^{\infty}d\omega^{E}\left[  \frac{\sin^{2}\left[  \left(  \omega
_{e}-\omega^{E}\right)  \tau\right]  }{\left(  \omega_{e}-\omega^{E}\right)
^{2}}\right]  \approxeq\int_{-\infty}^{\infty}d\omega^{E}\left[  \frac
{\sin^{2}\left[  \left(  \omega_{e}-\omega^{E}\right)  \tau\right]  }{\left(
\omega_{e}-\omega^{E}\right)  ^{2}}\right]  =\pi\tau,
\end{equation}
and%
\begin{equation}
\int_{0}^{\infty}d\omega^{E}\left[  \frac{\left(  1-\cos\left[  \left(
\omega_{e}-\omega^{E}\right)  \tau\right]  \right)  ^{2}}{\left(  \omega
_{e}-\omega^{E}\right)  ^{2}}\right]  \approxeq\int_{-\infty}^{\infty}%
d\omega^{E}\left[  \frac{\left(  1-\cos\left[  \left(  \omega_{e}-\omega
^{E}\right)  \tau\right]  \right)  ^{2}}{\left(  \omega_{e}-\omega^{E}\right)
^{2}}\right]  =\pi\tau.
\end{equation}

For the ground state where $\omega^{E}=\omega_{e}$, the integral in
$\omega^{E}$ collapses because of the resonance at $\omega^{E}=\omega_{e},$
and we have
\begin{equation}
P_{l,m}^{gain-E}=\frac{e^{2}}{M_{0}\gamma_{e}}\frac{\left(  \omega^{E}\right)
^{3}}{c^{3}}\left\langle \left[  J_{rad}\left(  \omega_{nl}^{E}\right)
\right]  \right\rangle \left\{  8\pi\frac{m^{4}\left[  Y_{lm}\right]  ^{2}%
}{l\left(  l+1\right)  }\left[  \frac{1}{x}\frac{d\left[  xj_{l}(x)\right]
}{dx}\right]  _{x=mv/c}^{2}\right\}  .
\end{equation}
This expression is to be compared to the power lost in the same radiation mode
given by Burko\cite{Burko}%
\begin{align}
&  P_{l,m}^{loss-E}=8\pi\frac{e^{2}}{c^{3}}m^{4}\omega_{e}^{4}r_{e}^{2}%
\frac{l\left(  l+1\right)  }{\left(  2l+1\right)  ^{2}}\left[  Y_{l,m}\left(
\pi/2,0\right)  \right]  ^{2}\left[  \frac{1}{l+1}j_{l+1}\left(  m\omega
_{e}r_{e}/c\right)  -\frac{1}{l}j_{l-1}\left(  m\omega_{e}r_{e}/c\right)
\right]  ^{2}\nonumber\\
&  =M_{0}c^{2}\left(  \frac{M_{0}c^{3}}{e^{2}}\right)  \frac{\left[
e^{2}/\left(  J_{\phi}c\right)  \right]  ^{8}}{1-\left[  e^{2}/\left(
J_{\phi}c\right)  \right]  ^{2}}\left\{  8\pi\frac{m^{4}\left[  Y_{l,m}\left(
\pi/2,0\right)  \right]  ^{2}}{l\left(  l+1\right)  }\left[  \frac{1}{x}%
\frac{d\left[  xj_{l}(x)\right]  }{dx}\right]  _{x=mv/c}^{2}\right\}
\end{align}
We obtain the condition for energy balance by equating the power lost and
gained,
\begin{equation}
P_{l,m}^{loss-E}=P_{l,m}^{gain-E}.
\end{equation}
\ Removing the common factors of
\begin{equation}
8\pi\frac{m^{4}\left\vert Y_{lm}\right\vert ^{2}}{l\left(  l+1\right)
}\left[  \frac{1}{x}\frac{d\left[  xj_{l}(mx)\right]  }{dx}\right]
_{x=v/c}^{2},\label{Common}%
\end{equation}
and simplifying, we find
\begin{equation}
J_{\phi}=J_{rad}=\hbar.\label{JeJrad}%
\end{equation}
In the ground state, the connection in Eq. (\ref{JeJrad}) gives average energy
balance between the loss of energy due to emission of radiation as dipole
radiation and the gain of energy from the dipole driving by zero-point
radiation for the relativistic charge. \ 

We expect that the ground state is completely stable, so that the energy
balance should hold not just for the dipole radiation, but for every radiation
mode. \ Burko\cite{Burko} gives the radiation emission for all modes when a
charged particle is moving in a circular orbit. \ \ We notice that the
variation from one mode to the next is contained in the factors which were
common to both the loss and gain of energy by the orbiting charge and are
given in the display (\ref{Common}). \ We expect this situation to continue
for the magnetic radiation modes.

\section{Resonant Excited States}

\subsection{Circular Resonant Excited States Have Lower Frequency}

The \textit{resonant excited states} of a charged particle in a Coulomb
potential are expected to be unstable and to decay with radiation emission
going down in energy to the stable ground state. \ Thus we will consider only
the average energy balance for \textit{dipole} radiation in the resonant
excited states. \ It is the absence of equilibrium for the higher multipole
which leads to the unstable behavior of the resonant excited states. \ On
radiation decay, it is usually the dipole radiation which is emitted as the
charged particle changes orbits from one excited state to a different state. \ 

For the circular resonant excited state labeled by $n$, the orbital angular
momentum $J_{e-n}$ is larger than that for the ground state $J_{e-1}.$
\begin{equation}
J_{e-n}=nJ_{e-1},
\end{equation}
as is the radius for the circular orbit
\begin{equation}
r_{e-n}=\left(  \frac{e^{2}}{M_{0}\gamma_{e-n}c^{2}}\right)  \left(
\frac{J_{e-n}c}{e^{2}}\right)  ^{2}=\left(  \frac{e^{2}}{M_{0}c^{2}}\right)
\sqrt{1-\left(  \frac{e^{2}}{J_{e-n}c}\right)  ^{2}}\left(  \frac{J_{e-n}%
c}{e^{2}}\right)  ^{2},
\end{equation}
while the orbital angular frequency is smaller%
\begin{equation}
\omega_{e-n}=\left(  \frac{M_{0}\gamma_{e-n}c^{3}}{e^{2}}\right)  \left(
\frac{e^{2}}{J_{e-n}c}\right)  ^{3}=\left(  \frac{M_{0}c^{3}}{e^{2}}\right)
\frac{1}{\sqrt{1-\left[  e^{2}/\left(  J_{e-n}c\right)  \right]  ^{2}}}\left(
\frac{e^{2}}{J_{e-n}c}\right)  ^{3}.
\end{equation}
These equations agree with those given for a circular orbit with the square
roots arising from the relativistic mass expression. \ %

\begin{equation}
P_{l,m}^{gain-E}=\frac{e^{2}}{M_{0}\gamma_{e}}\frac{\left(  \omega^{E}\right)
^{3}}{c^{3}}\left[  J_{rad}\left(  \omega_{nl}^{E}\right)  \right]  \left\{
8\pi\frac{m^{4}\left[  Y_{lm}\right]  ^{2}}{l\left(  l+1\right)  }\left[
\frac{1}{x}\frac{d\left[  xj_{l}(x)\right]  }{dx}\right]  _{x=mv/c}%
^{2}\right\}  \label{Pgain}%
\end{equation}%
\begin{align}
&  P_{l,m}^{loss-E}\nonumber\\
&  =M_{0}c^{2}\left(  \frac{M_{0}c^{3}}{e^{2}}\right)  \frac{\left[
e^{2}/\left(  J_{\phi}c\right)  \right]  ^{8}}{1-\left[  e^{2}/\left(
J_{\phi}c\right)  \right]  ^{2}}\left\{  8\pi\frac{m^{4}\left[  Y_{l,m}\left(
\pi/2,0\right)  \right]  ^{2}}{l\left(  l+1\right)  }\left[  \frac{1}{x}%
\frac{d\left[  xj_{l}(x)\right]  }{dx}\right]  _{x=mv/c}^{2}\right\}
\label{Ploss}%
\end{align}

Removing the common factors given in curly brackets in Eqs. (\ref{Pgain}) and
(\ref{Ploss}), the requirement for energy balance in a circular orbit is that%
\begin{equation}
\frac{e^{2}}{M_{0}\gamma_{e-n}}\frac{\left(  \omega_{e-1}/n^{3}\right)  ^{3}%
}{c^{3}}\left\langle \left[  J_{rad}\left(  \omega_{nl}^{E}\right)  \right]
\right\rangle =M_{0}c^{2}\left(  \frac{M_{0}c^{3}}{e^{2}}\right)
\frac{\left[  e^{2}/\left(  nJ_{e-1}c\right)  \right]  ^{8}}{1-\left[
e^{2}/\left(  nJ_{e-1}c\right)  \right]  ^{2}}.
\end{equation}
Simplifying, we find the energy-balance requirement is%
\begin{equation}
\frac{1}{nJ_{e-1}}J_{rad}=1. \label{1dnh}%
\end{equation}
The relationship implies that the zero-point radiation should be delivering to
the charge $n$ times as much power as being lost in the radiation emission.

\subsection{Higher Radiation Multipoles and Repeated Forces}

Because the orbit of the charge has finite extent, $r_{e-n}\neq0,$ it cannot
be treated as a point dipole. \ The charge density of the charge is given by
\begin{equation}
\rho_{e}(r,\phi,0,t)=e\delta^{2}\left[  \mathbf{r-}\widehat{x}r\cos\left(
\omega t\right)  -\widehat{y}r\sin\left(  \omega t\right)  \right]  ,
\end{equation}
which is a periodic function of time $t,$ but involves all the
\textit{multiples} of the fundamental orbital frequency\cite{Jackson4} \ Thus,
the charge density is given by
\begin{equation}
\rho_{e}(r,\phi,0,t)=e\sum\nolimits_{m=0}^{\infty}\rho_{m}(r)\cos\left[
m\omega_{e}t\right]
\end{equation}
where the Fourier transform is
\begin{equation}
\rho_{m}\left(  r\right)  =\frac{4\pi}{\omega_{e}}\int_{0}^{\omega_{e}/\left(
2\pi\right)  }dt\,\rho_{e}(r,\phi,0,t)\cos\left[  m\omega_{e}t\right]  .
\end{equation}

The zero-point radiation mode at frequency of the ground state, $\omega
_{rad}=\omega_{e-1},$ is going around faster that the slower orbital motion
$\omega_{e-n}=\omega_{e-1}/n^{3}$ at larger radius. \ Thus the rotating
radiation wave sweeps over the orbiting charged particle $n$ times, because
the zero-point radiation is still pushing the charged particle with the same
force and at the same frequency $\omega_{rad}=\omega_{n-1}=\omega_{e-1}$.
\ The charged particle's orbital frequency is much lower than when it is in
the ground state. \ For example, if $n=2,$ the speed of the charge in its
orbit is only half that of the ground state,%
\begin{equation}
r_{e-2}\omega_{e-2}=\left(  r_{e-1}n^{2}\right)  \left(  \omega_{e-1}%
/n^{3}\right)  =v_{e-2}=e^{2}/\left(  2J_{e-1}\right)  =e^{2}/\left(
2\hbar\right)  . \label{rewwe2}%
\end{equation}
This means that the wave will pass over the charge $n$ times while the charge
goes once around its orbit. \ But if the radiation wave now passes over the
particle $n$ times while the charge goes around its orbit once, then it picks
up $n$ times as much energy. \ This is exactly the factor of $n$ needed to
give energy balance for the \textit{dipole} radiation in the resonant excited
state in Eq. (\ref{1dnh}). \ 

\section{Nonrelativistic Limit}

\subsection{Large $c$ Limit}

In the limit as the speed of the charged particle is regarded as very small
compared to the speed of light $c,$ all the equations above go over to their
nonrelativistic limits. \ It seem fascinating that all the nonrelativistic
formulae of the Bohr theory are finite and do not involve the constant $c$.
\ Thus as $c$ is regraded as very large compared to any particle speeds, the
formulae of the relativistic expressions go over to the nonrelativistic
mechanical expressions where the constant $c$ does not enter. \ For example,
the particle speed in a circular orbit is simply
\begin{equation}
v_{e-n}=\frac{e^{2}}{n\hbar},\label{venr}%
\end{equation}
the nonrelativistic orbital radius is%
\begin{equation}
r_{e-n}=\frac{e^{2}}{M_{o}c^{2}}\left[  \sqrt{1-\left(  \frac{e^{2}}{J_{\phi
}c}\right)  ^{2}}\left(  \frac{J_{\phi}c}{e^{2}}\right)  ^{2}\right]
\rightarrow\frac{e^{2}}{M_{o}c^{2}}\left(  \frac{J_{\phi}c}{e^{2}}\right)
^{2}=\frac{n^{2}\hbar^{2}}{M_{0}e^{2}},\label{renBB}%
\end{equation}
\qquad the orbital frequency is
\begin{equation}
\omega_{e-n}=\frac{M_{0}c^{3}}{e^{2}}\left[  \frac{1}{\sqrt{1-\left[
e^{2}/\left(  J_{\phi}c\right)  \right]  ^{2}}}\left(  \frac{e^{2}}{J_{\phi}%
c}\right)  ^{3}\right]  \rightarrow\frac{M_{0}c^{3}}{e^{2}}\left(  \frac
{e^{2}}{J_{\phi}c}\right)  ^{3}=\frac{M_{0}e^{4}}{n^{3}\hbar^{3}%
}.\label{wenBB}%
\end{equation}
Bohr's theory\cite{Bohr} of 1913 involves exactly these same expressions
(\ref{venr}), (\ref{renBB}), and (\ref{wenBB}).

\subsection{Bohr Energy Levels for Hydrogen}

From the formulae (\ref{renBB}) and (\ref{wenBB}) we can obtain the energy in
the $n^{th}$ orbit as
\begin{equation}
U(J_{\phi})=\frac{1}{2}M_{0}v_{e}^{2}-\frac{e^{2}}{r_{e}}=\frac{1}{2}%
M_{0}\left(  \frac{e^{2}}{n\hbar}\right)  ^{2}-\frac{M_{0}e^{4}}{n^{2}%
\hbar^{2}}=-\frac{M_{0}e^{4}}{2n^{2}\hbar^{2}}\label{BB}%
\end{equation}
Here we have given a completely classical electromagnetic calculation of the
Bohr spectrum as the nonrelativistic result for a charged particle in a
Coulomb potential including zero-point radiation. \ We have not postulated the
integer values of angular momentum,$|\mathbf{L}|=n\hbar,$ but have derived the
integer values from the inclusion of Lorentz-invariant classical zero-point
radiation. \ 

Even when in a resonant excited state $U_{e-n}$, the radiative loss of energy
by the \textit{charge }in the Coulomb potential is mainly at the
\textit{dipole} frequency $\omega_{e-n}$. \ However, the energy gain for the
charge is due to the radiation at a higher frequency $\omega_{rad}=n\omega
_{e}$. \ The orbiting charge received $n$ impulses from the radiation which is
at the higher frequency. \ Thus, the charge's \textit{dipole} energy loss and
gain is balanced, even in the excited states. However, the radiation energy
loss is at frequency $\omega_{e-n}=\omega_{rad}/n,$ whereas the charge's
energy gain is at $n$ times the frequency $\omega_{e-n}$ of the orbiting
charge. \ The \textit{higher} \textit{driving multipoles} are, in general,
\textit{not} in energy balance; full energy balance for all the multipole
holds only when the oscillator is in its ground state. \ 

Radiation emission and absorption both involve the speed of light $c.$
\ However, radiation energy balance in zero-point radiation is independent of
the actual value for $c$, and the balance is hidden. \ We emphasize how
different this result is from the original Bohr postulate that a charged
particle in certain preferred orbits simply does \textit{not} radiate. \ Here
we find that the radiation is simply \textit{balanced} in certain resonant orbits.

\subsection{Energy Transitions}

The transition from one value of $n$ to another is associated with a change in
the radius of the charged particle orbit, and also in the frequency of the
charged particle radiation. \ If the integer $n$ changes by one unit from $n$
to $n-1$, then the change in average energy of the charged particle orbit is%
\begin{equation}
\Delta U_{e=n\rightarrow n-1}=-\frac{Me^{4}}{2\left(  n\hbar\right)  ^{2}%
}-\left(  -\frac{Me^{4}}{2\left[  \left(  n-1\right)  \hbar\right]  ^{2}%
}\right)  =\frac{Me^{4}}{2\left[  \left(  n-1\right)  \hbar\right]  ^{2}%
}-\frac{Me^{4}}{2\left(  n\hbar\right)  ^{2}}. \label{DUennm1}%
\end{equation}
From Eq. (\ref{wenBB}) and (\ref{DUennm1}), the transition energy is also
related to the frequency of the transition as the average of the energies of
the initial and final resonant states
\begin{equation}
\Delta U_{e=n\rightarrow n-1}=\frac{\left(  n-1\right)  }{2}\frac{M_{0}e^{4}%
}{\left(  n-1\right)  ^{3}\hbar^{3}}-\frac{n}{2}\frac{M_{0}e^{4}}{n^{3}%
\hbar^{3}}.
\end{equation}
In the limit of large resonant quantum number $n$, we have the Bohr
correspondence rule giving the emission frequency as essentially the same as
the frequency of the charge's orbital frequency
\begin{equation}
\Delta U_{e=n\rightarrow n-1}\approxeq\frac{Me^{4}}{2\left[  n\hbar\right]
^{2}}\left(  1+\frac{2}{n}+...\right)  -\frac{Me^{4}}{2\left(  n\hbar\right)
^{2}}\approxeq\frac{M_{0}e^{4}}{n^{3}\hbar^{3}},
\end{equation}
as given in Eq. (\ref{wenBB}).

In the situation of energy \textit{balance for the dipole radiation} at each
resonance excited state labeled by $n$, it may appear as though the charged
particle were not radiating at all, since the charge's average energy does not
change. \ Net radiation appears only on \textit{changes} of the index $n.$
\ During the transition, the charged particle radiation is not balanced by the
driving zero-point radiation. \ Once again, just as for the ground state, the
stabilizing role of the classical zero-point radiation in resonant excited
states may seem completely hidden. \ 

\section{Closing Summary}

In 1913 when Bohr suggested that in certain preferred orbits an electron did
not radiate, it occasioned profound skepticism among physicists. \ The
successes of classical electrodynamics were well known, and the emission of
radiation by an accelerating charged particle was widely accepted. \ However,
despite its unsettling ideas, Bohr's theory gave the correct results for the
wavelengths of the line spectra from hydrogen, even giving the correct reduced
mass correction for the positive helium ion. \ In 1916,
Sommerfeld\cite{Sommerfeld} extended Bohr's ideas to give the correct fine
structure of the hydrogen spectral lines. \ 

Today, the situation in physics is vastly different. \ Classical
electromagnetism is regarded as holding for macroscopic situations, but
quantum mechanics with its unusual ideas holds sway for the microscopic
domain. \ And as Feynman has claimed, \textquotedblleft I think I can safely
say that nobody understands quantum mechanics.\textquotedblright

In the present article, we suggest that Bohr's ideas of 1913 can be understood
in terms of classical electromagnetism with random classical zero-point
radiation. \ Although the idea of a \textit{classical zero-point radiation}
was considered by Nernst\cite{Nernst} in 1916, it never gained traction in the
physics community. \ The idea was treated seriously and extensively for the
harmonic oscillator by Marshall\cite{Marshall2} in 1963, and later by others.
\ But free fields or linear potentials gave the only successes. \ 

A summary article\cite{B1975} in 1975 attracted some attention. \ However, the
introduction of zero-point energy alone did not explain many phenomena which
seemed amenable to quantum treatment. \ Many physicists who were initially
enthusiastic about the idea of zero-point energy lost interest, and some
turned actively against it. \ A careful numerical hydrogen simulation by Cole
and Zou\cite{Cole2003} introduced a rare bright spot during this time. \ 

Apparently, it is not sufficient to introduce the idea of classical zero-point
radiation and to go on using nonrelativistic potentials for mechanical
systems. \ One must restrict attention to \textit{relativistic} or
\textit{approximately relativistic} systems. \ Also, Cole carried out
numerical calculations\cite{Cole2018} for hydrogen in the presence of a single
circularly polarized plane wave propagating perpendicular to the orbit, and he
noted the presence of resonances at multiples of the orbital frequency. \ In
the present work, we find the need to consider the \textit{resonances} between
the relativistic classical charged particle and the classical zero-point
radiation. \ When we have these three ingredients, \textit{classical
zero-point radiation, relativity, and resonance}, then classical
electrodynamics produces the same results for hydrogen that Bohr's original
old quantum theory proposed. \

\end{document}